
\magnification=\magstep1
\overfullrule=0pt
\setbox0=\hbox{{\cal W}}

\setbox9=\hbox{{\cal S}}

\def\w{{\cal W}}

\def\vac{\nu}
\def\lew{\lambda}

\def\undertext#1{$\underline{ \vphantom{y} \smash{ \hbox{#1} }}$}
\def \id{{\rm 1 \kern-2.8pt I }}

\def\lb{\lbrack}
\def\rb{\rbrack}

\def\q#1{\lb#1\rb}
\def\mn{\medskip\smallskip\noindent}
\def\sn{\smallskip\noindent}

\font\extra=cmss10 scaled \magstep0 \font\extras=cmss10 scaled 750

\setbox1 = \hbox{{{\extra R}}}
\setbox2 = \hbox{{{\extra I}}}
\setbox3 = \hbox{{{\extra C}}}

\setbox4=\hbox{{{\extra Z}}}
\setbox5=\hbox{{{\extras Z}}}
\setbox6=\hbox{{{\extras z}}}
\def\Z{{{\extra Z}}\hskip-\wd4\hskip 2.5 true pt{{\extra Z}}}

\def\BZT{{\rm Z{\hbox to 3pt{\hss\rm Z}}}}
\def\BZS{{\hbox{\sevenrm Z{\hbox to 2.3pt{\hss\sevenrm Z}}}}}
\def\BZSS{{\hbox{\fiverm Z{\hbox to 1.8pt{\hss\fiverm Z}}}}}
\def\BZ{{\mathchoice{\BZT}{\BZT}{\BZS}{\BZSS}}}
\def\BQT{\,\hbox{\hbox to -2.8pt{\vrule height 6.5pt width .2pt
    \hss}\rm Q}}
\def\BQS{\,\hbox{\hbox to -2.1pt{\vrule height 4.5pt width .2pt\hss}$
   \scriptstyle\rm Q$}}
\def\BQSS{\,\hbox{\hbox to -1.8pt{\vrule height 3pt width
   .2pt\hss}$\scriptscriptstyle \rm Q$}}
\def\BQ{{\mathchoice{\BQT}{\BQT}{\BQS}{\BQSS}}}
\def\BCT{\,\hbox{\hbox to -3pt{\vrule height 6.5pt width
     .2pt\hss}\rm C}}
\def\BCS{\,\hbox{\hbox to -2.2pt{\vrule height 4.5pt width .2pt\hss}$
   \scriptstyle\rm C$}}
\def\BCSS{\,\hbox{\hbox to -2pt{\vrule height 3.3pt width
   .2pt\hss}$\scriptscriptstyle \rm C$}}

\def\BHT{{\rm I{\hbox to 5.3pt{\hss\rm H}}}}
\def\BHS{{\hbox{\sevenrm I{\hbox to 4.2pt{\hss\sevenrm H}}}}}
\def\BHSS{{\hbox{\fiverm I{\hbox to 3.5pt{\hss\fiverm H}}}}}

\def\BPT{{\rm I{\hbox to 5pt{\hss\rm P}}}}
\def\BPS{{\hbox{\sevenrm I{\hbox to 4pt{\hss\sevenrm P}}}}}
\def\BPSS{{\hbox{\fiverm I{\hbox to 3pt{\hss\fiverm P}}}}}

\def\BST{\;\hbox{\hbox to -4.5pt{\vrule height 3pt width .2pt\hss}
   \raise 4pt\hbox to -2pt{\vrule height 3pt width .2pt\hss}\rm S}}
\def\BSS{\;\hbox{\hbox to -4.2pt{\vrule height 2.3pt width .2pt\hss}
   \raise 2.5pt\hbox to -4.8pt{\vrule height 2.3pt width .2pt\hss}
   $\scriptstyle\rm S$}}
\def\BSSS{\;\hbox{\hbox to -4.2pt{\vrule height 1.5pt width .2pt\hss}
   \raise 1.8pt\hbox to -4.8pt{\vrule height 1.5pt width .2pt\hss}
   $\scriptscriptstyle\rm S$}}

\def\BFT{{\rm I{\hbox to 5pt{\hss\rm F}}}}
\def\BFS{{\hbox{\sevenrm I{\hbox to 4pt{\hss\sevenrm F}}}}}
\def\BFSS{{\hbox{\fiverm I{\hbox to 3pt{\hss\fiverm F}}}}}

\def\BRT{{\rm I{\hbox to 5.5pt{\hss\rm R}}}}
\def\BRS{{\hbox{\sevenrm I{\hbox to 4.3pt{\hss\sevenrm R}}}}}
\def\BRSS{{\hbox{\fiverm I{\hbox to 3.35pt{\hss\fiverm R}}}}}

\def\BNT{{\rm I{\hbox to 5.5pt{\hss\rm N}}}}
\def\BNS{{\hbox{\sevenrm I{\hbox to 4.3pt{\hss\sevenrm N}}}}}
\def\BNSS{{\hbox{\fiverm I{\hbox to 3.35pt{\hss\fiverm N}}}}}
\def\BN{{\mathchoice{\BNT}{\BNT}{\BNS}{\BNSS}}}
\def\BAT{\hbox{\raise1.8pt\hbox{\sevenrm/}{\hbox to 4pt{\hss\rm A}}}}
\def\BAS{\hbox{\raise1.4pt\hbox{\fiverm/}
{\hbox to 3pt{\hss\sevenrm A}}}}
\def\BASS{\hbox{\raise1.4pt\hbox{\fiverm/}
{\hbox to 3pt{\hss\sevenrm A}}}}

\def\upin{\hbox{${\scriptstyle\cup}\hbox to
-2.7pt{\hss\vrule height 3.5pt}$}}
\def\downin{\hbox{${\scriptstyle\cap}\hbox to
-2.7pt{\hss\vrule height3.5pt}$}}

\def\Sequenz #1,#2,#3{0\longrightarrow
#1\longrightarrow #2\longrightarrow #3
   \longrightarrow 0}

\def\teiltnicht{\mathrel
{\raise1pt\hbox to .5pt{$\scriptstyle/$\hss}\vert}}
\def\r{{\cal R}}
\def\fk{{\cal N}}
\def\bpz{1}
\def\bai{2}
\def\blg{3}
\def\bal{4}
\def\bou{5}
\def\blm{6}
\def\www{7}
\def\ver{8}
\def\cas{9}
\def\mor{10}
\def\fuc{11}
\def\nob{12}
\def\koh{13}
\def\mfl{14}
\def\dor{15}
\def\mat{16}
\def\eho{17}
\def\ehd{18}
\def\nil{19}
\def\pkm{20}
\def\sch{21}
\font\HUGE=cmbx12 scaled \magstep4
\font\Huge=cmbx10 scaled \magstep4
\font\Large=cmr12 scaled \magstep3

\font\large=cmr17 scaled \magstep0
%
%
\nopagenumbers
\pageno = 0
\centerline{\HUGE Universit\"at Bonn}
\vskip 10pt
\centerline{\Huge Physikalisches Institut}
\vskip 2.5cm
\centerline{\Large Fusion Algebras Induced by }
\vskip 6pt
\centerline{\Large Representations of the Modular Group }
\vskip 1.6cm
\centerline{\large W.\ Eholzer}
\vskip 1.5cm
\centerline{\bf Abstract}
\vskip 15pt
\noindent
Using the representation theory of the subgroups $SL_2(\BZ_p)$ of
the modular group we investigate the induced fusion algebras in
some simple examples. Only some of these representations lead to
'good' fusion algebras.
Furthermore, the conformal dimensions
and the central charge of the corresponding rational conformal field
theories are calculated. Two series of representations which can
be realized by unitary theories are presented.
We show that most of the
fusion algebras induced by admissible representations
are realized in well known rational models.
\vfill
\settabs \+&  \hskip 110mm & \phantom{XXXXXXXXXXX} & \cr
\+ & Post address:                       & BONN-HE-92-30   & \cr
\+ & Nu{\ss}allee 12                     & hep-th/9210040  & \cr
\+ & W-5300 Bonn 1                       & Bonn University & \cr
\+ & Germany                             & October 1992    & \cr
\+ & e-mail:                             & ISSN-0172-8733  & \cr
\+ & eholzer@mpim-bonn.mpg.de            &                 & \cr
\eject
\pageno=1
\footline{\hss\tenrm\folio\hss}
\leftline{\undertext{\bf 1. Introduction }}
\mn
Starting with the seminal work of A.A. Belavin, A.M. Polyakov and
A.M. Zamolodchikov in 1984 $\q{\bpz}$ there have been several
Ans\"atze to classify rational conformal field theories (RCFT).
In one approach extensions of the Virasoro algebra are
obtained either by the free field construction starting from
Kac-Moody algebras  $\q{\bai-\bal}$ or by other direct explicit methods
(see $\q{\bou,\blm}$ and references therein).
For extensions of the Virasoro algebra
the modular properties of the highest weight representations and
their fusion algebras have been investigated $\q{\www,\ver}$.
In another approach one deals with abstract fusion algebras and
tries to classify them $\q{\cas}$.
\mn
In this paper we exclusively deal with
fusion algebras induced by representations of the modular group where
we call a fusion algebra {\it induced} if it can be calculated
from a representation of the modular group using the Verlinde formula
having chosen a vacuum state in the representation space.
The aim of this approach is the classification of all
physically relevant representations of the modular group and thus
a classification of RCFTs.
So far it seems that all representations which satisfy some
natural criteria for "good" fusion rules are realized in RCFT.
\mn
This paper is organized as follows.
In chapter two we will discuss fundamental properties of RCFT which
are important for later discussions.
In chapter three  we give a short review of the representations of
special subgoups of the modular group. Then we investigate
these representations in detail and show that most of the physically
relevant representations are realized by some RCFT.
Finally we summarize our results and point out some open questions.
\mn\mn
\leftline{\undertext{\bf 2. Fundamental properties of RCFT}}
\mn
Let $\r$ be a rational conformal field theory with (extended)
symmetry  algebra $\w$ such that $\w$ contains the Virasoro algebra
as a subalgebra.
The finite set of $\w$-primary fields $\{ \phi_i \}_{i=1}^n$ of $\r$
correspond to the highest weight representations ${\cal H}_i$
of $\w$. Let $h_i$ be the  conformal dimensions of the primary
fields $\phi_i$ and
denote the vacuum representation by ${\cal H}_{\vac}$ with
$h_{\vac} = 0$.
The conformal characters of the representations ${\cal H}_i$ read
$$ \chi_i(\tau) := tr_{{\cal H}_i}(q^{L_0 - {c\over24}})
   \ \ \ \ {\rm with} \ \ \ \ q :=e^{2 \pi i \tau}     \eqno{(2.1)}$$
where $c$ is the central charge of the theory and $\tau$ is the
modular parameter. These conformal characters form
a finite dimensional (right)
 representation $r$ of the modular group $SL_2(\Z)$.
This group is generated by the two transformations
$$ T = \pmatrix{ 1 &1  \cr 0 &1 \cr} \ \ {\rm{ and }} \ \
   S = \pmatrix{ 0 &-1 \cr 1 &0}  \eqno{(2.2)} $$
and acts on the characters as
$$ \bigl( r(A) \chi_i \bigr) (\tau) :=
   \chi_i( {\textstyle {a\tau+b\over c\tau+d}} )
   \ \ \ \    \forall A= \pmatrix{a &b \cr c &d}  \in SL_2(\Z)
   \eqno{(2.3)} $$
The modular transformation $T$ is represented
by a diagonal matrix with nonvanishing elements
and $S^2$ by the identity matrix, if
one chooses the conformal characters $(2.1)$
as a basis of the representation space.
Note that $r(S)^2= \id$ is only true if one considers the
characters as $q$-dimensions of the representations ${\cal H}_i$
(if the characters additionally take  account of charge quantum
numbers, $r(S)^2$ equals the 'conjugation' in the
fusion algebra which is a matrix of order 2 (see e.g. $\q{\cas}$).
\mn
Since we are considering a rational theory  the dimensions $h_i$
of the primary fields and the central charge $c$ are rational $\q{\mor}$.
This implies the existence of a positive integer $m$ such that
$$ r(T^m) = \id.  \eqno{(2.4)} $$
If the kernel of the representation $r$ of the modular group
contains a principal congruence subgroup $\Gamma_m$
it factorizes to a representation of
$SL_2(\Z_m)$ and $(2.4)$ is evident.
This is true for all explicitly known examples
so that we will restrict ourselves to representations of
$SL_2(\Z_m)$.
\mn
If the representation $r$ is known explicitly one can calculate the
fusion algebra
$$ \phi_i \times \phi_j = \sum_{k=1}^n  \fk_{ij}^k \phi_k.  \eqno{(2.5)}$$
using the famous Verlinde formula $\q{\ver}$
$$ \fk(\vac)_{ij}^k = \sum_{m=1}^n
       { r(S)_{i,m} r(S)_{j,m} r(S^{-1})_{m,k} \over
                                r(S)_{\vac,m} } \eqno{(2.6)} $$
where we have written the dependence of the fusion coefficients on the
vacuum state explicitly.
In order to be able to interpret the fusion coefficients
$\fk(\vac)_{ij}^k$ as dimensions
of the corresponding intertwiner spaces these coefficients should be
nonnegative integers. This is what we call a 'good' fusion algebra.
\mn
Assume that the conformal dimensions $h_i$ of the primary fields
are nondegenerate  ($i \ne j \Rightarrow h_i - h_j \notin \Z$).
Denote the lowest energy representation by ${\cal H}_{\lew}$
($i \ne \lew \Rightarrow h_{\lew} < h_i $), which in the unitary
case is the vacuum representation.
Then the quantum dimensions of the
representations ${\cal H}_i$ with respect to ${\cal H}_{\lew}$ are given by
$$ \delta_k = \lim_{q \to 1}{ \chi_k(q) \over \chi_{\lew}(q)}=
              \lim_{\tau \to 0}{ \chi_k(\tau) \over \chi_{\lew}(\tau)}=
              \lim_{\tau \to i\infty}{ \chi_k(-{1 \over \tau})
                    \over \chi_{\lew}(-{1 \over\tau})}=
              \lim_{\tau \to i\infty}{ \bigl( r(S) \chi_k \bigr)(\tau)
                    \over \bigl( r(S) \chi_{\lew} \bigr) (\tau)}=
              {r(S)_{\lew,k} \over r(S)_{\lew,\lew} }. \eqno{(2.7)} $$
We have used that the conformal characters are dominated in the large
$ Im(\tau)$ limit by the leading term $q^{h_i-{c\over 24}}$ and that the
main contribution is given by the lowest energy state.
Since the quantum dimensions describe the relative dimensions of the
representations ${\cal H}_i$ they are positive reals.
For unitary theories one knows (see e.g. $\q{\fuc}$) that for
$$ \delta_k < 2 \ \ \Rightarrow  \ \
   \delta_k \in \bigl\{ 2 cos({\textstyle {\pi \over j+2}} )
                \bigr\}_{j \in \BN}.
   \eqno{(2.8)} $$
\mn
Furthermore, the quantum dimensions $\delta_k$ satisfy the fusion algebra
$$ \delta_i \delta_j = \sum_{k=1}^n \fk_{ij}^k \delta_k.   \eqno{(2.9)} $$
\mn
Starting from this setup one possible way to classify RCFT arises
from the representation theory of $SL_2(\Z_m)$: given a representation
of $SL_2(\Z_m)$, what RCFT corresponds to it, if any ?
In particular, the following questions occur:
\mn
\item{$\bullet$} Is it possible to identify the lowest
                 energy representation such that the resulting
                 quantum dimensions are positive reals and lead
                 for the case of unitary theories to the ones
                 given by $(2.8)$ ?
\item{$\bullet$} Is there an element of the representation space
                 corresponding to the vacuum,  such that the fusion
                 coefficients defined by $(2.6)$ are positive integers ?
\mn
The irreducible representations of $SL_2(\Z_m)$ can be obtained
using Weil representations (c.f. $\q{\nob}$).
A Weil representation is a representation of $SL_2(\BZ)$ in the space
of complex-valued functions over a finite $\BZ_m$ module. In these
representations $T$ is represented by a pointwise multiplication
and $S$ by a discrete Fourier transformation. For more details as
well as for the mathematical definition of Weil representations we
refer to $\q{\nob}$.
Because it is always possible to represent $r(T)$ unitarily one can choose
a basis such that $r(T)$ becomes diagonal.
\mn
If for a special representation the answers to these two questions
are positive we will call the representation "conformally admissible".
{}From now on this kind of representations will just be called
admissible.
\mn
In the following we will concentrate to the case of
irreducible representations of $SL_2(\Z_m)$.
Though there are well known RCFT which
lead to reducible representations of the modular group
a reduction to the irreducible case might be possible.
For example, RCFT with nondegenerate conformal dimensions $h$ lead
to irreducible representations of the modular group.
Even some representations with degenerate conformal dimensions
like e.g.\ the
Virasoro minimal models corresponding to nondiagonal partition
functions contain irreducible admissible subrepresentations.
The restriction to irreducible representations is in this case
justified because  the subrepresentations
are realized on $\w$-algebra characters $\q{\blm,\eho}$.
Similarly, for particular fractional level Kac-Moody algebras with
reducible representations
(cf. the example in $\q{\koh}$) one irreducible part is
admissible and realizes a Virasoro minimal model.
\mn
The irreducible representations of $SL_2(\Z_m)$ are known.
To construct them it is sufficient to know the irreducible
representations of $SL_2(\Z_p)$ where $p$ is a prime power.
Since the investigation of fusion algebras induced by representations
of $SL_2(\BZ_p)$ ($p$ an odd prime) is very simple for representations
with nondegenerate conformal dimensions
we will further restrict on this case in the next chapters.
\mn\mn
\leftline{\undertext{\bf 3. Irreducible representations of
                            $SL_2(\Z_p)$ with $p$ an odd prime }}
\mn
The irreducible representations of $SL_2(\Z_p)$ can be classified
by their characters (characters are understood in the group theoretical
 sense,
and have to be distinguished from the conformal characters $(2.1)$).
Using the notation of L. Dornhoff $\q{\dor}$
there are the following inequivalent representations:
\mn
\centerline{table 1: {\it the irreducible representations of
$SL_2(\BZ_p)$}}
\sn
\centerline{
\vbox{ \offinterlineskip
\def\tablespace{ height2pt&\omit&&\omit&&\omit&&\omit&&
                           \omit&&\omit&&\omit&\cr }
\def\tablerule{ \tablespace
                \noalign{\hrule}
                \tablespace      }
\hrule
\halign{&\vrule#&
  \strut\quad\hfil#\hfil\quad\cr
\tablespace
& representation && $\id$ && $\psi$ && $\chi_i$    && $\theta_j$
                 && $\xi_1,\xi_2$           && $\eta_1,\eta_2$       &\cr
\tablerule
& dimension      && 1     && $p$    && $p+1$       && $p-1$
                 && ${1\over2}(p+1)$        && ${1\over2}(p-1)$      &\cr
\tablerule
& $r(S^2)$       && $\id$ && $\id$  && $(-1)^i\id$ && $(-1)^j\id$
                 && $(-1)^{p-1\over2}\id$   && $(-1)^{p+1\over2}\id$ &\cr
\tablespace}
\hrule}
}
$$ {\rm with} \ \ \ 1 \le i \le {1\over2}(p-3) \ \ ,
    \ \ 1 \le j \le {1\over2}(p-1).  \eqno{(3.1)} $$
To avoid confusion we tolerate by convention that
both the conformal characters and
certain irreducible representations of $SL_2(\BZ_p)$
have been denoted by $\chi_i$.
The table shows that there are altogether $p+4$ inequivalent
irreducible representations of $SL_2(\Z_p)$.
We have also listed representations with $r(S^2) \ne \id$ since
we will consider in the sequel equivalence classes of
representations which differ only by a one dimensional representation
of the modular group. These one dimensional representations can
be realized by powers of the Dedekind eta function (for a detailed
discussion see chapter 4).
\mn
The trivial representation $\id$ is of no real interest since the
corresponding quantum dimension and fusion algebra is trivial.
We will now consider some of the nontrivial representations in general
and verify that most of
the admissible representations are realized in some RCFT.
To this end we have worked out the explicit form of $r(T)$ and $r(S)$
where
$r$ is one of the representations in the table. This is done by using
quadratic modules and the Weil representations (for the general outline
see $\q{\nob}$).
The explicit form of $r(T)$ shows for all representations except
$\chi_i$ that in the corresponding RCFT the conformal dimensions
$h_j$ cannot be degenerate, because the diagonal elements of $r(T)$
are pairwise different. The particular form of
the matrices $r(T)$ and $r(S)$ for the representations
is given below.
\mn
Because $T$ should be represented by a diagonal matrix and $S$ by a
unitary  one
the only remaining freedom in the choice of a basis
in the nondegenerate case is a
conjugation with a diagonal matrix whose entries are roots of unity.
If one fixes a possible lowest energy state $(\lew)$ of the
representation space corresponding to ${\cal H}_{\lew}$ with
$$ r(S)_{k,\lew} \ne 0  \ \ \ \ 1 \le k \le n,   \eqno{(3.2)} $$
the basis is completely determined by the requirement of
positive real quantum dimensions (condition $(3.2)$ is
necessary in order to give well defined quantum dimension (c.f.\ $(2.5)$)).
The set of all $(\lew)$ which obey $(3.2)$ will be denoted by ${\cal F}_r$.
In the next step one can fix another element $(\vac)$ of the
representation space corresponding to the vacuum representation
${\cal H}_{\vac}$ and ask whether this choice leads to "good"
fusion rules defined by the Verlinde formula.
This is the program we will follow  for some simple examples
in the next chaper.
\mn\mn
\leftline{\undertext{\bf 4. Explicit results }}
\mn
We now investigate which irreducible representations of $SL_2(\Z_p)$,
where $p$ is an odd prime, are admissible.
\mn
\leftline{\undertext{The representations $\eta_{1,2}$}}
\mn
Consider the representations $\eta_{1,2}$. The dimensions of the
representation spaces are $n = {1\over2}(p-1)$ and the representations
are given by the matrices
$$ \eqalign{
     r(T)_{k,k} &=  e^{{2 \pi i\over p} a k^2}  \cr
     r(S)_{k,j} &=  {2 i \over \sqrt{p}} \bigl({a\over p}\bigr)
                    \epsilon(p) sin \bigl({2\pi\over p}2 a k j\bigr) \cr
              1 &\le k,j \le n
   }\eqno{(4.1)}  $$
$$ {\rm where} \ \ \epsilon (p) = \cases{ 1, &$p \equiv 1 \  (mod \ 4)$ \cr
                                          i, &$p \equiv 3 \  (mod \ 4)$ \cr
   } $$
and $\bigl({a\over p}\bigr)$ is the Legendre symbol
($\bigl({a\over p}\bigr)$ is 1 if a is a square in $\Z_p$ and -1
otherwise).
The two different representations are obtained for two values of $a$
namely $\eta_1$ if $a$ is a square and $\eta_2$ if $a$ is no square in
$\Z_p$.
Note that for $p$ odd but not prime and $a \in \Z_p$ invertible, eq.\
$(4.1)$ also defines a representation of $SL_2(\Z_p)$.
\mn
The choice of an  element $(\lew)$ which obeys $(3.2)$ is free, but fixes
already the element $(\vac)$, if one
requires the fusion coefficients to be positive integers.
If the pair $(\lew,\vac)$ leads to nonnegative fusion
coefficients the exchanged pair $(\vac,\lew)$ also gives
positive fusion coefficients.
The corresponding conformal dimensions $hi$ for a given vacuum
state $(\vac)$ are obtained from $(4.1)$
$$  h_k \equiv {a\over p}(k^2-\vac^2)  \ \ (mod \ \BZ).  \eqno{(4.2)} $$
Using the well known relation between the central charge $c$ and the
conformal dimensions $h_i$ the central charge is given by $\q{\mat}$
$$ c \in {24\over n}\sum_{i=1}^n h_i -2(n-1) + {4\over n}
     \bigl( \BN\setminus\{1\} \bigr)
     =  3 - p -{24a\over p}\vac^2 +{48\over p-1}\BZ +
     {4\over n} \bigl( \BN\setminus\{1\}  \bigr)
   \eqno{(4.3)} $$
\mn
In the following table we list the possible pairs $(\lew,\vac)$ for
$p=3,5$ and $7$:
\mn
\centerline{table 2: {\it possible lowest energy and vacuum states}}
\centerline{{\it for the representations $\eta_{1,2}$ and $p \le 7$}}
\sn
\centerline{
\vbox{ \offinterlineskip
\def\tablespace{ height2pt&\omit&&\omit&&\omit&\cr }
\def\tablerule{ \tablespace
                \noalign{\hrule}
                \tablespace      }
\hrule
\halign{&\vrule#&
  \strut\quad\hfil#\hfil\quad\cr
\tablespace
& $p$             && $\eta_1$             && $\eta_2$             &\cr
\tablerule
& $3$             && $(1,1)$              && $(1,1)$              &\cr
\tablerule
& $5$             && $(1,1),(2,2)$        && $(1,2),(2,1)$        &\cr
\tablerule
& $7$             && $(1,2),(2,1),(3,3)$  && $(1,3),(2,2),(3,1)$  &\cr
\tablespace}
\hrule}
}
\mn
where we choose $a=1$ to define $\eta_1$ and
$a = min \{ m \in \BN \vert \ \bigl({m\over p}\bigr) = -1 \}$
to define $\eta_2$.
\mn
A representation can only be realized by a unitary RCFT if
the states $(\lew)$ and $(\vac)$ are equal, so that for example
for $p=5$ the representation $\eta_2$ cannot by realized in any unitary
RCFT.
For the above admissible representations which could be
realized by unitary RCFT the quantum dimensions are
\mn
$$ \eqalign{
   p&=3:  \ \ \ \ \delta^{\eta_{1,2}}_1 = 1 = 2cos({\pi\over3})  \cr
   p&=5:  \ \ \ \ \delta^{\eta_{1,2}}_i \in
                               \{ 1,2 cos \bigl( {\pi\over5} \bigr) \}\cr
   p&=7:  \ \ \ \ \delta^{\eta_{1,2}}_i \in
             \{ 1,2 cos \bigl( {\pi\over7} \bigr),\beta \} \cr
   {\rm and} \ 2 < \beta & \
   {\rm is \ the \ largest \ positive \ root \ of}
    \ x^3-2x^2+1.
  }\eqno{(4.4)} $$
This implies that all of the above representation with
$(\lew)$ =  $(\vac)$ can be realized by a unitary theory.
\mn
Looking for candidates of unitary theories one finds
for every $p$ exactly two possible ones.
These series are given by $(4.1)$ with $a=1$ and $a=-1$ for
$\lew = \vac ={1\over2}(p-1)$. It is easy to calculate
the corresponding quantum dimensions
$$ \delta_{{1\over2}(p-1)} = 1 \ \ \ \ , \ \ \ \
     \delta_1= 2 cos \bigl( {\pi\over p} \bigr) \ \ \ \ , \ \ \ \
     \delta_k > 2 \ \ \ {\rm for} \ \ \ \ 1 < k < {1\over2}(p-1).
   \eqno{(4.5)} $$
\mn
The cases $p=3,5,7$ are in complete agreement with the results of
$\q{\cas}$ and the possible fusion algebras
in the above table lead exactly to the $h$- and
$c$-values given in the classification of fusion algebras corresponding
to RCFT with one, two and three conformal characters in $\q{\cas}$.
\mn
There are two remarkable facts concerning these fusion algebras.
First, the fusion algebras induced by $\eta_1$ and $\eta_2$ are
equivalent.
This is seen using the simple recurrence relation
$$ \fk(\vac)_{\widehat{i+\vac},\widehat{j+\vac} }^k =
                              \fk(\vac) _{i,j}^k +
                              \delta_{0,i+j+\vac-k}-
                              \delta_{0,i+j+\vac+k} \eqno{(4.6a)}$$
with
$$\eqalign{
   \widehat{k} &= \cases{ k,  &$ k \le n$ \cr
                           p-k,&else} \cr
   \delta_{a,b} &= \cases{ 1, &$ a \equiv b \ (mod \ p)$ \cr
                           0, &else }. \cr
   }\eqno{(4.6b)}$$
which follows from standard  trigonomentric identities.
This relation is independent of $a$ so that the representations
$\eta_1$ and $\eta_2$ induce the same fusion algebra.
\mn
Secondly, one finds that  all fusion algebras
for different elements  $(\vac)$ of the representations
$\eta_{1,2}$ are equivalent
$$  \fk(1)_{i,j}^k = \pm \fk(\vac)_{\widehat{\vac \cdot i}\ ,
    \ \widehat{\vac\cdot j}}^{\widehat{\vac \cdot k}}
    \ \ \forall \vac \in {\cal F}_{\eta_{1,2}}
    \eqno{(4.7)}$$
where ${\cal F}_{\eta_{1,2}}$ is given by
$$ \{ a \ \vert \ \ 1 \le a \le n \ \ , \ \ (a,p)=1 \ \} \eqno{(4.8)} $$
which clearly equals $M_n = \{1,\ldots,n\}$ if $p$ is a prime.
\mn
Similar symmetry properties will occur for most of the
representations we consider.
\mn
We have seen that the representations $\eta_{1,2}$ are indeed
admissible, but:  are they realized by a RCFT ?
Looking closer at $r(S)$ and $r(T)$ in $(4.1)$ one observes that for
$2 a \equiv 1 \ (mod \ p)$ these matrices are up to a phase
equal to the ones for the
Virasoro minimal models with central charge $c=c(2,p)$.
The phase can be understood in the following way.
It is always possible to write the conformal characters as
$$ \chi_j = \eta^{-\alpha} \Lambda_j   \eqno{(4.9)}  $$
where $\eta$ is the Dedekind eta function and the $\Lambda_j$ are
cusp forms of weight ${\alpha \over 2}\in \BN$ if $\alpha$
is large enough.
There are two possible natural choices for $\alpha$. Either
$\alpha$ is chosen minimal or such that the representation
of the modular group on the $\Lambda_j$ factorizes to a
representation of $SL_2(\BZ_m)$ with minimal $m$.
In both cases the representation $r$ in the space of conformal characters
factorizes into a one-dimensional one acting on $\eta^{-\alpha}$ and
an $n$-dimensional one acting on the modular
forms $\Lambda_j$.
Writing the conformal characters of the Virasoro minimal models $c(2,p)$
as in $(4.9)$ with
$$\alpha = \alpha(p) = \cases{ 4, &$p \equiv 7 \ (mod \ 8)$ \cr
                              10, &$p \equiv 5 \ (mod \ 8)$ \cr
                              16, &$p \equiv 3 \ (mod \ 8)$ \cr
                              22, &$p \equiv 1 \ (mod \ 8)$ \cr
  } \eqno{(4.10)} $$
leads to a representation on the $\Lambda_j$ given by $(4.1)$ with
$2 a \equiv 1 \ (mod \ p)$. This generalizes:
even for $p$ odd but not prime and $2a \equiv 1 \ (mod \ p)$
$(4.1)$  leads to the correct $r(S)$
and $r(T)$ matrices for the Virasoro minimal models with
central charge $c=c(2,p)$.
Using the notation of $\q{\cas}$ this implies that the representations
$\eta_{1,2}$ of $SL_2(\BZ_p)$ induce fusion algebras of type
${\cal B}_{{p-3\over2}}$.
\mn
For $p=5$ and $a=1,-1$ there are
two possible candidates for unitary RCFT as can be seen from  table 2.
These two are realized by the level 1 WZW models corresponding
to $G_2$ and $F_4$ with $\alpha = 22$ and $10$, respectively. Both
theories lead to the same representations of the modular group
but their $h$-values differ (see eq. $(4.2)$).
We have not yet found a realization for the other members of the
two unitary series.
\mn\mn
\leftline{\undertext{The representations $\xi_{1,2}$ }}
\mn
The dimension of the representation spaces is $n={1\over2}(p+1)$ and
$r(T)$ and $r(S)$ are given by
$$ \eqalign{
   r(T)_{k,k} &=  e^{{2\pi i\over p}a k^2}  \cr
   r(S)_{k,l} &=  {2\over \sqrt{p}} \epsilon(p)
   \bigl({a\over p} \bigr) cos \bigl( {2\pi\over p}2a k l \bigr)
    \ \ \  1\le k,l \le n-1 \cr
   r(S)_{0,l} &= r(S)_{l,0} =
             {2\over \sqrt{p}} \bigl( {a\over p} \bigr) \epsilon(p)
              \ \ \ l \ne 0\cr
   r(S)_{0,0} &= {1\over \sqrt{p}} \bigl( {a\over p} \bigr) \epsilon(p).
              \cr
   }\eqno{(4.11)} $$
In complete analogy to $\eta_{1,2}$ one obtains the two different
representations $\xi_{1,2}$ with two values of $a$ namely $\xi_1$ with
$\bigl({a\over p} \bigr) = 1$ and $\xi_2$ with
$\bigl({a\over p} \bigr) = -1$. Furthermore $(4.11)$ also gives
representations for nonprime odd $p$.
\mn
Using standard trigonometric identities one obtains for the fusion
coefficient with respect to $\vac = 0$
$$ \eqalign{
   \fk(0)_{0,j}^k &= \delta_{j k}  \cr
   \fk(0)_{i,j}^k &= \cases{ {1\over \sqrt{2}},
                             &$ i \pm j \pm k \equiv 0 \ (mod \ p)$ \cr
                             0, & otherwise } \ \
   \rm{ if} \ i,j,k \ne 0.
   }\eqno{(4.12)}$$
Note that these equations are independent of $a$.
For $\vac \ne 0 $ one obtains
$$ \fk(\vac)_{i j}^k \in \{0,\pm 1,\pm {1\over \sqrt{2}}, \pm \sqrt{2} \}
   \eqno{(4.13)} $$
and
$$ \fk(1)_{i j}^k =
   \pm \fk(\vac)_{\widehat{\vac i},\widehat{\vac j}}
                ^{\widehat{\vac k}} \ \ \ \forall \vac \ne 0.
   \eqno{(4.14)}$$
\mn
The calculation of the fusion algebras shows that
this series of representations cannot be realized by RCFT.
\mn
One can show, however, that  direct sums of two representations
$\xi$ exist such that the resulting fusion algebra is integer valued.
In order to ensure the fusion coefficients to be integers one has to
use two representations for odd numbers $p$ and $\tilde p$ with
$p - \tilde{ p} = 4$ and perform a unitary change of basis mixing
the two states labelled by $1$.
These fusion algebras are realized by the parabolic
$\w(2,8k)$ algebras with $4k$ odd $\q{\mfl}$.
\mn\mn
\leftline{\undertext{The representations $\theta_j$ }}
\mn
To give the explicit form of the matrices $r(T)$ and $r(S)$ we have to
fix some notation.
Let $\BFT_p$ the field
obtained from $\Z_p$ by quadratic extension,
$\epsilon$ be a generator of the cyclic
group $\BFT^*_p$ and $\rho_j = e^{2 \pi i {j \over p+1}}$.
Using this notation the matrices $r(T)$ and $r(S)$ are given by
$$ \eqalign{
    r(T)_{k,k} &= e^{ {2\pi i \over p} \epsilon^{k}
                             \overline{\epsilon^{k}} }  \cr
    r(S)_{k,l} &= {1\over p} \sum_{m=0}^{p} \rho_j^m
                  e^{ {2\pi i \over p}
                  2 Re( \epsilon^{(p-1)m} \epsilon^{k}
                                  \overline{\epsilon^{l}}) } \cr
               0 &\le k,l \le p-2.
   } \eqno{(4.15)} $$
The inequivalent representations $\theta_j$ are obtained for
$1 \le j \le {1\over2}(p-1)$.
Calculating the fusion coefficients for these representations for
$p=3,5,7,11$ leads to admissible representations
 only if $ p \equiv 5 \ (mod \ 6) $ and
$j = {1\over 3}(p+1)$.
Continuing this series one finds admissible representations
for $p=17$ and $23$.
As in the case of the representations $\eta_{1,2}$ the choice
of an element $(\lew)$ already fixes the state $(\vac)$.
Since these states are always
different, these representations cannot be realized by a unitary RCFT.
The corresponding $h$- and $c$-values can be read off from $(4.15)$
using $(4.3)$.
Furthermore, all fusion algebras obtained
for different elements $(\vac)$ from these representations
are equivalent, since
$$ \eqalign{
    \fk(1)_{i,j}^k &= \pm
    \fk(\vac)_{\sigma_{\vac}(i) \sigma_{\vac}(j)}^{\sigma_{\vac}(k)}  \cr
    \sigma_{\vac}(k) &= k+\vac \ (mod \ p-1).
   } \eqno{(4.16)}$$
\mn
In contrast to the representations
$\eta_{1,2}$ there seems to be no canonical extension of these
four examples since calculations show that for $23 < p \le 167$
the resulting fusion algebras are not integer valued.
\mn
The case $p=3$ corresponds to a theory with a single character. These
theories have been classified completely in ref. $\q{\sch}$.
For $p=5$ one obtains that
$\theta_2 = \eta_1 \otimes \eta_2$ such that this representation
is realized in at least two RCFT.
It is  known that the representations with  $p=11,17,23$ are
realized by the exceptional
$\w(2,\delta)$ algebras with $\delta=4,6,8$ and
$c_{eff} = { 4(\delta -1) \over 3\delta -1}$ $\q{\eho,\ehd,\nil}$.
For the realization of the conformal characters of the corresponding
$\w(2,\delta)$ algebras as well as for more details we
refer to $\q{\nil}$.
The explicit calculations of the fusion algebras for $p=5,11,17,23$ show
that the maximal fusion coefficients for the corresponding
representations are $1,3,6,12$.
\mn
Since the other representations $\theta_j$ for $p=3,5,7,11$ and
all $\theta_j$ for $p=29$ are not admissible, we expect that
the ones mentioned above are the only admissible representations.
\mn\mn
\leftline{\undertext{The representation $\psi$ }}
\mn
The dimension of the representation space is $n=p$ and the
representation is given by
$$ \eqalign{
   r(T)_{k,k} &= e^{{2\pi i\over p} k}  \cr
   r(S)_{k,l} &= {1\over p} \sum_{a=1}^{p-1}
                 e^{{2 \pi i \over p}(a k + a^{-1} l )}
                 \ \ \ \ 1 \le k,l \le p-1  \cr
  r(S)_{k,0} &= r(S)_{0,k} = {\sqrt{p+1} \over p} \ \ \ k \ne 0 \cr
  r(S)_{0,0} &= -{1\over p}.
 } \eqno{(4.17)} $$
\mn
A simple calculation leads to  the fusion
coefficients with respect to $\vac = 0$
$$ \eqalign{
   \fk(0)_{0,j}^k &= \delta_{j k}  \cr
   \fk(0)_{i,j}^k &= {1 \over \sqrt{p+1}}( f_{i,j}^k - 1 )
   \ \ {\rm if} \ i,j,k \ne 0 \cr
   {\rm with} \ \ f_{k,l}^m &=
   {1\over p}\bigl(
                 \sum_{a,b,c\in \Z_p^* \atop a+b+c \equiv 0 \ (mod \ p)}
                      e^{{2 \pi i}(a^{-1}k+b^{-1}l+c^{-1}m)} \ \
                    -2  \bigr) \cr
             &\in \{0,\pm 1 \}.
   }\eqno{(4.18)}$$
For $\vac = 1$ one obtains
$$ \fk(1)_{i j}^k \in \BQ({1\over\sqrt{p+1}})
   \eqno{(4.19)} $$
so that the smallest field containing the fusion coefficients is
$\BQ({1\over\sqrt{p+1}})$.
Furthermore we have
$$ \fk(1)_{i j}^k =
   \pm \fk(\vac)_{{\widehat \vac i},{\widehat \vac j}}
                ^{{\widehat \vac k}} \ \ \ \forall \vac \ne 0.
   \eqno{(4.20)}$$
\mn
This shows that there are no RCFT corresponding to
the representation $\psi$.
\mn\mn
\leftline{\undertext{The representations $\chi_l$ }}
\mn
Let $b$ be a generator of the cyclic group $\Z^*_p$ and
$\rho_l=e^{2\pi i {l\over p-1}}$.
With this notation the matrices $r(T)$ and $r(S)$ for the representations
$\chi_l$ read
$$\eqalign{
  r(T)_{j,j} &=  e^{{2 \pi i\over p}j}   \cr
  r(S)_{j,k} &= {1\over p} \sum_{m=1}^{p-1} \rho_l^m
                e^{{2\pi i \over p}(b^m j+b^{-m} k )}
                \ \ {\rm for} \ \  0\le j,k \le p-1 \cr
  r(S)_{p,k} &= (-1)^l r(S)^*_{k,p} =
        {1\over p}\sum_{m=1}^{p-1}\rho_l^m e^{{2\pi i\over p}b^{-m}}
        \ \ {\rm for} \ \ 0\le k \le p-1 \cr
  r(S)_{p,p} &= 0.
  }\eqno{(4.21)}$$
\mn
The diagonal elements of $r(T)$ are not pairwise different, so that
corresponding RCFT must have degenerate $h$-values.
We have not investigated this case in detail, but
we found no admissible representations for $p=5,7,11$.
\mn
\mn\mn
We should remark that the new symmetry property
$(4.7),(4.14),(4.16),(4.20)$  arises not only in these
four examples but also for fractional level
Kac-Moody algebras (cf. the example presented
in $\q{\koh}$) and for parabolic $\w$-algebras $\q{\pkm}$.
\mn\mn\mn
\leftline{\undertext{\bf 5. Summary and open questions }}
\mn
We have presented a new approach to the classification of
RCFT starting from irreducible representations of the modular group.
For the case of $SL_2(\Z_p)$ with $p$ an odd prime we determined
the possible physically relevant representations and found
that most of these cases are realized by well-known RCFT.
Furthermore, we observed a new symmetry property of fusion algebras
induced by representations of the modular group.
\mn
In the future more general investigations of physically relevant
representations might lead to a better understanding
of the representations of the modular group and RCFT.
In particular, it should be possible to derive a classification of
all representations corresponding to
RCFT with nondegenerate conformal dimensions.
In this context it would be interesting to understand the connection
between known RCFT and reducible representations of the modular group.
\mn\mn
\leftline{\undertext{\bf Acknowledgements}}
\mn
I would like to stress that without the constant advise  of
N. Skoruppa concerning the representation theory of the modular
group this work would not have been possible.
\mn
I am indebted to the Max-Planck-Institut f\"ur Mathematik
in Bonn-Beuel because all calculations were performed on their
computers.
\mn
Of great help were the remarks of M. Flohr, M. Terhoeven and
R. Blumenhagen concerning parabolic $\w$-algebras and
(fractional level) Kac-Moody algebras.
\mn
I am grateful to M. Terhoeven for careful reading of the manuscript.
\mn
Finally I would like to thank
A. Honecker, R. H\"ubel,
J. Kellendonk, S. Mallwitz, W. Nahm, A. Recknagel,
M. R{\"o}sgen and R. Varnhagen for many helpful discussions.

\mn\mn
\leftline{\undertext{\bf References}}
\mn
\settabs\+\indent&$\q{1000}$\quad&\cr
\+$\q{\bpz}$ & A.A. Belavin, A.M. Polyakov, A.B. Zamolodchikov \cr
\+         & {\it Infinite Conformal Symmetry in Two-Dimensional
   Quantum Field Theory}  \cr
\+         & Nucl. Phys. {\bf B}241 (1984) p. 333  \cr
\+$\q{\bai}$ & F.A. Bais, P. Bouwknegt, M. Surridge, K. Schoutens \cr
\+         & {\it Coset Construction for Extended Virasoro Algebras} \cr
\+         & Nucl. Phys. {\bf B}304 (1988) p. 371 \cr
\+$\q{\blg}$ & A. Bilal, J.L. Gervais \cr
\+         & {\it Systematic Construction of Conformal Theories with
   Higher-Spin Virasoro Symmetries} \cr
\+         & Nucl. Phys. {\bf B}318 (1989) p. 579 \cr
\+$\q{\bal}$ & J. Balog, L. Feh\'er, P. Forg\'acs, L. O'Raifeartaigh,
   A. Wipf \cr
\+         & {\it Kac-Moody Realization of $\w$-Algebras} \cr
\+         & Phys. Lett. {\bf B}244 (1990) p. 435 \cr
\+$\q{\bou}$ & P. Bouwknegt, {\it Extended Conformal Algebras},
 Phys. Lett. {\bf B}207 (1988) p. 295 \cr
\+$\q{\blm}$ & R. Blumenhagen, M. Flohr, A. Kliem, W. Nahm,
A. Recknagel, R. Varnhagen \cr
\+         & {\it $\w$-Algebras with Two and Three Generators},
 Nucl. Phys. {\bf B}361 (1991) p. 255 \cr
\+$\q{\www}$ & W. Nahm, {\it A Proof of Modular Invariance }\cr
\+         & Int. Jour. Mod. Phys. {\bf A}6 (1991) p. 2837 \cr
\+$\q{\ver}$ & E. Verlinde, {\it Fusion Rules and Modular
Transformations in 2D Conformal Field Theory} \cr
\+         & Nucl. Phys. {\bf B}300 (1988) p. 360\cr
\+$\q{\cas}$ & M. Caselle, G. Ponzano, F. Ravanini, {\it Towards
a Classification of Fusion }\cr
\+          & {\it Rule Algebras in Rational Conformal Field Theories}\cr
\+          & Int. J. Mod. Phys. {\bf B}6      (1992) p. 2075 \cr
\+$\q{\mor}$ & G. Anderson, G. Moore, {\it Rationality in
Conformal Field Theory}\cr
\+           & Comm. Math. Phys. 117 (1988) p. 441 \cr
\+$\q{\fuc}$ & J. Fuchs, {\it Quantum Dimensions}, preprint
CERN-TH.6156/91 \cr
\+$\q{\nob}$ & A. Nobs, J. Wolfart, {\it Die irreduziblen
Darstellungen der Gruppen $SL_2(Z_p)$,}  \cr
\+         & {\it insbesondere $SL_2(Z_2)$  I \& II } \cr
\+         & Comment. Math. Helvetici 39($\bf51$) p. 465-489,
p. 491-526 \cr
\+$\q{\koh}$ & I.G. Koh, P. Sorba, {\it Fusion Rules and
(Sub-)Modular Invariant }\cr
\+          & {\it Partition Functions in Non-Unitary Theories }\cr
\+          & Phys. Lett. {\bf B}215 (1988) p. 723 \cr
\+$\q{\mfl}$ & M. Flohr, {\it $\w$-Algebras, New Rational Models
and Completeness }\cr
\+          & {\it of the $c=1$ Classification}, preprint BONN-HE-92-08 \cr
\+$\q{\dor}$ & L. Dornhoff, {\it Group representation theory},
Marcel Dekker Inc., New York (1971)\cr
\+$\q{\mat}$ & D. Mathur, S. Mukhi, A. Sen, {\it On the Classification
of Rational Conformal Field Theories } \cr
\+           & Phys. Lett. {\bf B}213 (1988) p. 303 \cr
\+$\q{\eho}$ & W. Eholzer, M. Flohr , A. Honecker , R. H{\"u}bel,
W. Nahm, R. Varnhagen  \cr
\+         & {\it Representations of $\w$-Algebras with Two Generators
and New Rational Models } \cr
\+         & preprint BONN-HE-91-22 \cr
\+$\q{\ehd}$ & W. Eholzer, {\it Exzeptionelle und Supersymmetrische
$\w$-Algebren }\cr
\+          & {\it in Konformer Quantenfeldtheorie}, Diplomarbeit
BONN-IR-92-10 \cr
\+$\q{\nil}$ & W. Eholzer, N. Skoruppa, {\it Exceptional $\w$-Algebra
Characters and }\cr
\+          & {\it Theta-Series of Quaternion Algebras, }
              in preperation  \cr
\+$\q{\pkm}$ & private communication with M. Flohr \cr
\+$\q{\sch}$ & A. N. Schellekens, {\it Meromorphic $c=24$
Conformal Field Theories}\cr
\+           & preprint CERN-TH.6478/92 \cr
\vfill

\end